\documentstyle[openbib]{elsart}
\def\msun{$M_\odot \,$}
\def\mdot{$\dot M\,$}

\def\sles{\lower2pt\hbox{$\buildrel {\scriptstyle <}
   \over {\scriptstyle\sim}$}}
\def\sgreat{\lower2pt\hbox{$\buildrel {\scriptstyle >}
   \over {\scriptstyle\sim}$}}

\begin{document}
\begin{frontmatter}
\title{Towards a Synthesis of Core-Collapse\\ Supernova Theory}
\author{Adam Burrows\thanksref{NSF}}
\address{Departments of Physics and Astronomy\\
University of Arizona\\
Tucson, Arizona, USA 85721\\}
\thanks[NSF]{Supported in part by the U. S. National Science 
Foundation}
\maketitle

\begin{abstract}

New insights into the mechanism and character of core--collapse 
supernova explosions
are transforming the approach of theorists to their subject.  The 
universal realization that the direct hydrodynamic
mechanism does not work and that a variety of hydrodynamic 
instabilities can influence
the viability of theoretical explosions
has ushered in a new era in supernova modeling.  
In this paper, I discuss the important physical and technical issues
that remain.
I review the neutrino--driven mechanism, the possible roles of 
Rayleigh--Taylor
instabilities, questions in neutrino transport, and the various
observational constraints within which theorists must operate. 
However, a consensus has yet to be achieved among active workers 
concerning many important details and some essential phenomenology.  
This synopsis  
is meant to accomplish two things: 1) to focus attention on the 
interesting problems
whose resolution will bring needed progress, and 2) to assess the 
current status of 
the theoretical art.

\end{abstract}
\end{frontmatter}
\bigskip

\section{Introduction}

A new synthesis is emerging in the theory of core-collapse supernovae.
This is not meant to imply that the basic mechanism has been found in
their multi-dimensional character or that a compelling consensus has 
been amicably reached.  Rather, new information and new ideas are 
accumulating 
at such a rapid rate on both the theoretical and observational fronts
that questions and facts long ignored by ``collapse'' theorists can no
longer be relegated to an indefinite future.  The 
epiphany of SN1987A was a turning point, but new computational 
capabilities
that emerged in the interim have also played a role, as has the 
exponentiation of astronomical data in the eighties and nineties.  
A complete
theory of supernova explosions must explain a variety of observations 
and facts.  Among the questions of relevance are:

\begin{enumerate}

\item What determines the scale of the supernova explosion energy 
(E$_{SN}$)?
Why are E$_{SN}$ for SN1987A and SN1993J $\sim 1.5 \pm 0.5 \times 
10^{51}$ergs \cite{tnh96}?

\item What is the mass spectrum of the residual neutron stars and 
pulsars?
Why is the average mass of the well-measured pulsars $\, 
\sim\!1.35 M_{\odot}$
\cite{thor93}?

\item Which progenitors leave black holes?  What is their mass 
spectrum?
What determines the masses of the X-ray Nova primaries, observed 
to range from
\sgreat 3.0 \msun to 16 \msun \cite{vpm95}?

\item What are the radioactive \nuc{56}{Ni}, \nuc{57}{Ni}, 
\nuc{26}{Al}, \nuc{44}{Ti},
\nuc{60}{Fe} yields as a function of progenitor mass?  Why are the
$\left[^{57}Fe \over ^{56}Fe\right]$ and $\left[^{44}Ca \over 
^{56}Fe\right]$ ratios what
they are and what does this imply for the explosion mechanism 
and mass cut?

\item Similarly, what is the dependence of the iron-peak 
({\it e.g.}, \nuc{54,56,58}{Fe},
\nuc{60,61,62}{Ni}, \nuc{59}{Co}, \nuc{55}{Mn}, \nuc{45}{Sc}, 
\nuc{52}{Cr},
\nuc{48}{Ti}, \nuc{64,66,68}{Zn}, \nuc{51}{V}, \nuc{63,65}{Cu}) 
yields on progenitor star
and mechanism?

\item When, how, and where are the r-process nuclei produced and 
ejected? Do different
progenitors have different r-process yields?

\item What is the explanation for the intermediate mass, iron-peak, 
and r-process
abundances observed in low-metallicity halo stars 
\cite{mcwill,mcwill2,sned94}?

\item Is there a range of intrinsic kicks or recoils imparted to 
the nascent neutron star that
can explain the high proper motions observed \cite{hla93}?  Is 
there a bimodal
distribution of natal kicks and, if so, why \cite{no90}?  Is 
there a correlation
between progenitor mass and kick magnitude and/or duplicity?  
What can be learned
about the supernova explosion from pulsar spin-orbit angles 
\cite{wcc95}?

\item Do neutrinos play a significant role in the synthesis of 
\nuc{11}{B}, \nuc{19}{F}, etc.
\cite{whhh90}?

\item What caused the heterogeneity and asphericity of the debris 
clouds in SN1987A, SN1993J,
Cas A, N132D, and the Crab \cite{fma,macalpine,lawr}?  Are they 
a consequence predominantly 
of core or mantle instabilities?

\item Is the ${\Delta Y} \over{\Delta Z}$ of galactic chemical 
evolution connected with
the creation of stellar mass black holes \cite{m92,psc94}?

\item What do the abundance patterns and high metallicity in QSO 
absorption line systems and Lyman--$\alpha$
clouds at high redshifts say about the evolution and explosion of 
massive stars?

\item What determines the magnetic fields of radio and X-ray pulsars?
With what multipolarity structure are pulsars born and why?

\item What is the spin period of pulsars at birth?  Is it 2 
milliseconds or
200 milliseconds? Is there a correlation between birth period, 
progenitor duplicity,
and nascent magnetic field?

\end{enumerate}

The mapping between progenitor mass and supernova characteristics 
(energy, mass cut, yields,
recoils, B--fields, {\it etc.}) is the crucial goal of theoretical 
supernova research.
There is a solid, but evolving, literature on the relationship 
between progenitor mass,
metallicity, and mass loss history and the density, entropy, and 
electron fraction
structures of the progenitor cores that eventually implode 
\cite{wlw95}.  Modern supernova 
theory seems posed to explain the influence of core structures 
on mass cut, post-bounce delay to
explosion, explosion energy, and nickel and r--process, {\it etc.} 
yields, but
has yet to do so.  Theoretical questions that will first need to be 
addressed and
answered are:

\begin{enumerate}

\item Is mass fallback a generic feature of supernova explosions?

\item How does progenitor structure affect the delay to explosion, 
explosion energy,
nucleosynthetic yields, and residual neutron star mass?

\item Is convection exterior to the gain radius (Mayle \cite{m85}) 
crucial, useful, or merely of
secondary importance to the explosion?

\item What is the role of deeper core instabilities (salt--finger, 
semi--convection,
or lepton--driven) in the explosion and in the  subsequent 
evolution of the 
protoneutron star?

\item What is the post--bounce delay to explosion?

\item How will more accurate neutrino transport influence the 
mechanism and 
development of the explosion?

\item How will a more accurate treatment of the neutrino opacities 
at high densities
alter the neutrino light curves and spectra and, hence, the 
core--mantle
coupling so crucial to the explosion?

\item What are the remaining important issues surrounding the 
nuclear equation
of state?

\item What is the influence, if any, of neutrino viscosity on the 
character and
nature of the explosion?

\item  Does rotation play an important role in the supernova explosion 
or in 
supernova observables?

\item Does the asphericity and heterogeneity of the explosion 
itself influence 
the mixing of \nuc{56}{Ni} into the outer stellar mantle, as observed 
in SN1987A?

\item When does the protoneutron star wind that follows explosion turn 
on and upon 
what physics does its emergence depend?

\item Can this wind be the site of the r--process?  Which progenitor 
structures
are more likely to yield r--process elements at infinity?

\item What neutrino physics determines the entropy and $Y_{\rm e}$ 
of this wind?

\item Can black hole formation be accompanied by a supernova 
explosion?

\item Are pulsar ``kicks'' in part a consequence of aspherical mass 
motions and/or
neutrino emissions that accompany collapse, bounce, and explosion?  
What are the 
essential hydrodynamic ingredients? Is there a progenitor dependence?

\item Do neutrino oscillations or exotic neutrino physics play a role?

\item Do hydrodynamic instabilities at birth influence pulsar 
B--fields?

\item What can one learn about the supernova mechanism from the 
detection of a
galactic neutrino burst by SuperKamiokande, SNO, or any one of the 
Gran Sasso detectors
(LVD, MACRO, Borexino, ICARUS)?

\item What can be learned about the internal dynamics of ``collapse'' 
from
its gravitational radiation signature in LIGO \cite{a92} or VIRGO 
\cite{ag92}?

\end{enumerate}

The program implied by the above two lists of questions is daunting and 
will
require the collective efforts of theorists and observers over many 
years.
My purpose in assembling these lists is to focus theorists in 
particular
on the variety of observational constraints to which their models must
already conform and to suggest interesting, but neglected, theoretical 
topics
that might be profitably explored.  In this paper, I summarize or 
provide a 
commentary on a few of the most interesting of these facts, ideas, and
outstanding issues.  In \S 2, I describe the basics 
of the supernova explosion phenomenon.
In \S 3, I summarize and analyze the effects of hydrodynamic overturn. 
Section 4 addresses
new issues concerning neutrino transport and opacities.  Section 5 
discusses
a few interesting constraints imposed by nucleosynthesis and the 
r--process.
Various mechanisms of pulsar natal kicks 
are touched on in \S 6, which includes a short discussion of the 
gravitational
wave signature of one of the kick mechanisms.  In \S 7, I wrap up 
with a few general comments.
Throughout, I assume that the
reader is familiar with both the technical issues and the history of 
the subject.
This allows me to concentrate on the interesting questions on the 
frontier
of supernova theory, while avoiding diversionary minutiae.

\section{Perspectives on the Mechanism}

It is now generally accepted that the prompt shock \cite{cj60} stalls
at a radius between 80 and 150 kilometers, due to photodissociation, 
neutrino losses,
and the accretion ram \cite{bl85,bc90,bhf95,bruenn85,bruenn89a}.  
The focus of supernova theory
is now on the subsequent behavior of this quasi-steady accretion shock 
on 
timescales of tens of milliseconds to seconds.  The essence of 
a supernova
explosion is the transfer of energy from the core to the mantle.  
The mantle
is less bound than the core, whose binding energy can grow without 
penalty
during the delay to explosion.  The core is the protoneutron star 
that will
evolve due to neutrino cooling and deleptonization over many seconds 
\cite{bl86}.
Bethe \& Wilson \cite{bw85} showed that it is possible and plausible 
that neutrino
heating of the accreted material near the shock could, over time, 
lead to an 
explosion.  That neutrinos mediate this energy transfer and are the 
agents of explosion
seems compelling \cite{b87,janka93,bethe90,bethe93a,bethe93b}.  
However, this said, the roles of many 
important phenomena and processes still need to be delineated and
refined.  Foremost among these are the multi-dimensional hydrodynamic 
effects
\cite{bhf95,hbhfc94,jm96,bethe95} and the neutrino opacites that 
regulate the
driving neutrino luminosities.  Before I address those topics, 
a few words on
the explosion condition itself may prove useful.  

If there were no post-bounce accretion of the outer mantle, it can be 
shown that
due to neutrino heating the material exterior to the inner core 
would be
{\bf unstable} to outflow.  This follows directly from simple 
arguments, analogous
to those employed by Parker \cite{Park} in his early ruminations 
concerning
the solar wind, that the atmosphere above the neutrinosphere can not 
be simultaneously
in hydrostatic and ``thermal'' equilibrium.  In the solar wind 
context, Parker
showed that, since the thermal conductivity of the plasma is almost 
independent
of density and depends on the temperature to a stiff power 
($\propto T^{5/2}$),
the equilibrium temperature profile would be a shallow function of 
radius ($\propto r^{-2/7}$).
He then went on to show that since the power $2/7$ was less than $1$, 
hydrostatic 
equilibrium of an ideal gas mantle would require a finite pressure 
at infinity.
Without such a pressure, a wind would be driven by electron 
conduction heating.
While this is no longer a viable model for the solar wind itself, 
these physical
arguments translate directly to the neutrino heating/cooling 
context of the 
post--bounce protoneutron star.  Since the neutrino heating rate per 
baryon goes
as 1/r$^2$ and the neutrino cooling rate per baryon goes as T$^6$, 
the equilibrium
temperature profile in a static atmosphere would go as 1/r$^{1/3}$.  
Since $1/3$ 
is below $1$, a finite pressure at large distances is required 
in order to 
thwart the spontaneous excitation of a vigorous outflow.  
Protoneutron star
atmospheres are unstable to neutrino-driven ejection.  
It is continuing accretion
(and the large pressures it affords) that suppresses the 
``explosion'' of the mantle.
However, as is clear from hydrodynamic simulations, the accretion 
rate need not 
decay to nothing before the mantle lifts off.  Deriving the precise 
time and
circumstances of the transition from quasi-steady accretion to 
explosion
requires some sophistication.  The presence, due to accretion, 
of more neutrino-energy-absorbing mass than would be available in the 
thin wind context
ultimately results in a more energetic explosion \cite{bethe95}.  
However, the outflow
will always eventually make the transition to a thin stable wind 
\cite{b87,bhf95,qian,dsw86}.

Another way to look at the condition for explosion is to note that, 
in order to eject matter,
neutrino heating must result in a steady matter temperature in the 
shocked mantle
that is above the ``escape temperature,'' crudely derived by setting 
the specific 
internal energy equal to $GM/r$.  This is the so-called ``coronal'' 
condition, familiar
in many other astronomical contexts.  Because heating occurs 
predominantly behind
a shock stalled at finite radius ($R_s$), there may not be enough 
matter or volume
exterior to the gain radius that satisfies the coronal condition 
and the mantle will
not explode.  Anything that enlarges the region in which the coronal 
condition is
satisfied pushes the object closer to the supernova threshold.  
Hence, it is 
advantageous to increase $R_s$.  Two agencies that can do this are 
an increased
neutrino luminosity and overturning convection near the shock 
\cite{hbc92,bethe90,bhf95,jm96}.

To discover the 
critical conditions for explosion and its subsequent development 
requires a full
hydrodynamic code, but insight into the pre-supernova structure and 
its stability can be
gained by studying it as a quasi-static structure in equilibrium 
\cite{bg93}.  This allows
one to convert the partial differential equations of hydrodynamics 
into simple
ordinary differential equations and the problem into an eigenvalue 
problem.  For
a given core mass, equation of state, neutrino transport algorithm, 
neutrino
luminosity ($L_\nu$), and mass accretion rate (\mdot), the shock 
radius (the eigenvalue)
and the stellar profiles (the eigenfunctions) can be derived.  
For a given core mass, 
$R_s$ as a function of the control parameters $L_\nu$ and \mdot 
can be obtained.
What Burrows \& Goshy \cite{bg93} showed was that for a given \mdot, 
there was
a critical $L_\nu$ above which there was no solution for $R_s$.  
They identified 
this critical $L_\nu$ versus \mdot curve with the approximate 
condition for the 
onset of explosion.  After the expansion commences, the problem 
must be handled
hydrodynamically.  However, expansion decreases the matter temperature 
and, hence, the
cooling rate faster than it decreases the heating rate and the 
instability should
run away for a given core $L_\nu$.  Since expansion cuts accretion 
and accretion
contributes in part to $L_\nu$, a concommitant decrease in $L_\nu$ 
might be of 
concern.  However, there is a time delay in the decrease of $L_\nu$ 
due to
expansion equal to the matter settling time from the shock to the 
core.  This can 
be a comfortable $\sim$30 milliseconds and is larger for larger 
pre-explosion
shock radii.  Again, anything that increases $R_s$ brings the star 
closer to the
explosion threshold.  

The calculations and assumptions of Burrows \& Goshy \cite{bg93}
were crude, particularly in the transport sector, but illuminate 
semi-quantitatively
the basics of the phenomenon.  A similar analysis for quite a 
different system,
AM Her objects, was performed by Chevalier \& Imamura \cite{ci82} 
and those who
have difficulty understanding Burrows \& Goshy \cite{bg93} are 
heartily referred
to that paper.  Note that it is the essence of equilibrium that 
timescales
of the relevant processes are comparable:  hydrostatic equilibrium 
is ``equivalent''
to the equality of sound--travel and free--fall times.  In the 
quasi-static
pre-explosion phase of the protoneutron star bounded by an 
accretion shock,
equilibrium is equivalent to the equality of the heating/cooling 
time ($\tau_\nu$)
and the settling time ($\tau_s$) of matter as it sinks from the 
shock to the core
(\sles $R_s/v$).  The shock radius, as the eigenvalue of the problem,
adjusts to accomplish this equality of timescales.  Hence, that 
$\tau_\nu$ and 
$\tau_s$ are comparable is a requirement of equilibrium, and is not 
a problem of the 
quasi-static assumption.  It is only when the characteristic 
timescale for the
{\it change} of \mdot becomes comparable to the other relevant 
times that the 
quasi-static assumption is dubious.  Indeed, that timescale can at 
times approach
$\tau_\nu$ and $\tau_s$, but is often significantly longer.  When it 
is short, the
core plus shock must be handled hydrodynamically.  However, when it is 
long, the
core plus mantle can also be subjected to a pulsation (perturbation) 
analysis
around an equilibrium structure.  In direct analogy with the standard 
stellar
pulsation problem, there are driving regions due to neutrino heating, 
damping regions
due to neutrino cooling and damping due to shock motion \cite{bm82}.  
A mode 
stability analysis of a protoneutron star bounded by an accretion 
shock may prove
illuminating.


\section{The Role of Multi-Dimensional ``Convective'' Motions}

That in the protoneutron star and supernova contexts there
should be hydrodynamic instabitilites (Rayleigh--Taylor, salt--finger,
semi--convection) has been known and studied since the work
of Epstein \cite{e79}.  A review of this literature can be found
in \cite[hereafter BHF]{bhf95}.  However, the role of convection and 
overturn has been controversial and ambiguous from the outset.
Many, evoking Ockam's Ravor, have opted to ignore it.  This
should no longer be possible.  There are three classes of 
instabilities
to address:  those in the core near and below the neutrinospheres,
overturning and boiling motions due to heating from below between the
gain radius and the shock \cite{bethe90,hbc92,bhf95,jm96}, and
Rayleigh--Taylor and Richtmyer--Meshkov instabilities in the outer
stellar mantle far beyond the ``iron'' core.  In this paper, I will 
ignore
the latter and concentrate on the former, since they are more 
germane to the explosion mechanism.

Convection below and near the neutrinospheres has been invoked to
boost the driving neutrino luminosities.  Mayle \& Wilson 
\cite{mw88,w85}
suggested that ``neutron--fingers'', akin to salt--fingers 
in the Earth's
oceans, advect sufficient heat outward to enhance the emergent 
neutrino
luminosities by twenty or more percent and, thereby, to turn a fizzle
into an explosion.  Without such a boost, they did not obtain 
explosions
and handled this convection with a mixing--length prescription.
Bruenn \& Dineva \cite{bd96} have recently challenged the
existence of such a finger instability in this context with a 
compelling
analysis of the details of energy and lepton transport in the 
protoneutron
star core.  Burrows \cite{b87} suggested that standard lepton-- or
entropy--driven convection beneath the neutrinospheres could provide a 
similar boost and BHF do obtain such an enhancement during the 
hundreds
of milliseconds after bounce.  However, in those calculations, 
while such
convective motions are definitely present, it is difficult to 
disentangle
this effect from everything else going on.  An enhancement of 
5\% to 20\% may
be inferred, but the jury is still out on the magnitude and 
importance of such
convective motions, driven in part by negative lepton gradients 
maintained
as the protoneutron star deleptonizes from without.

It is neutrino--driven convection (overturn) near the stalled shock 
that
has recently achieved prominence.  Following the suggestion by Bethe 
\cite{bethe90},
Herant {\it et al.} \cite[hereafter HBC]{hbc92} 
\cite[hereafter HBHFC]{hbhfc94}, BHF,
and Janka \& M\"uller \cite[hereafter JM]{jm96} demonstrated the 
positive effect of such
convective motions in aiding (perhaps enabling) the supernova 
explosion.
However, the different groups interpret the specific role of these 
multi--dimensional motions differently.  Here, I will briefly 
lay out the
issues, but with the obvious bias.

The stalling shock dynamically creates a negative entropy gradient that
is unstable to overturn on slightly longer timescales (5--15 
milliseconds).
This convection is the first multi--dimensional effect of note 
after bounce
\cite{bf92}.  However, as shown by BHF, Wilson \cite{mw88}, and Bruenn 
\cite{bruenn89b},
neutrino transport smooths out this gradient within 25 milliseconds.  
Importantly,
as demonstrated by BHF and HBC another maximum in
entropy develops at the gain radius, exterior to
which the matter is unstable to overturn, in a matter akin to the 
boiling
of water on a stove.  A new convective zone is established between 
the gain radius and
the shock, but the mantle does not yet explode.  Instead, the 
protoneutron star,
now with a convection zone, evolves quasi--statically for fifty to 
hundreds of
milliseconds as it continues to accrete matter through the shock.  
BHF claim
that most of the accreted matter eventually settles onto the core 
and does not 
dwell more than a few convective cycles in the gain region.  
However, while it
does cycle it has longer to absorb neutrino energy.  The net effect 
is a larger
average entropy in the convective gain region than can be attained 
in one--dimension.
In 1--D, matter moves through the gain region quickly and deliberately 
\cite{hbc92}.
Entropies of 10--20 are achieved.  In the two--dimensional 
calculations of BHF,
average entropies in the gain region reach 25--35.  At these entropies,
the matter is {\it not} radiation--dominated and the term 
``hot bubble''
is inappropriate.  Since convective motions
smooth out the average entropy distribution in the gain region, and 
the 
average entropies themselves are larger, the entropies near the 
shock are
larger than in 1--D.  Atmospheres with larger entropies have larger 
radii,
all else being equal.  These larger radii are just what the discussion 
in \S 2
claimed were advantageous for explosion and are 
the major consequence of
multi--dimensional effects.  However, this does not guarantee an 
explosion.
It merely facilitates it by lowering the explosion threshold. 
In sum, the
critical luminosity is lower in 2--(3--)D than in 1--D, but neutrinos 
are still 
in the driver's seat.

Not all workers agree with this description.  
Bethe in particular claims 
that soon 
after the establishment of the convective zone, matter and energy 
``accumulate''
in the gain region and matter does not leak onto the core (see also 
HBC).  The
matter accreted through the shock and its energy ``build'' until the 
total energy
in the gain region, including that due to nuclear recombination, 
reaches
approximately 10$^{51}$ ergs, at which time the mantle explodes.  
Bethe points
out that overshoot in three dimensions is weaker than in two 
dimensions,
so that the leak of matter onto the core and out of the gain 
region might
be plugged.  This must be explored.  Nevertheless, a few of the 
ingredients of 
the paradigm are problematic.  First, neither BHF nor JM
see such a building, until the explosion commences.  The mass and 
energy in the
region actually decrease prior to the instability.  In  addition, 
in the 
calculations of BHF, the net total enthalpy (read energy) flux is 
inwards, not
outwards, despite the rising plumes.  Such plumes must contend 
with inward
accretion and are balanced by downward moving plumes.  Second, 
due to electron
capture, the 0.1--0.2 \msun that would reside in the gain region 
would be
very neutron--rich and, to be consistent with severe nucleosynthetic 
constraints,
must not be ejected.  This requires that a lot of matter must fall 
back during the
latter stages of the explosion.  While this is not implausible for 
the most bound 
stellar progenitors (\sgreat\, 20 \msun) \cite{bhf95,ww96}, it is 
difficult to
understand for the lighter progenitors (9--15 \msun), whose binding 
energies are
quite modest, but which dominate the IMF.  Furthermore, without 
fallback the gravitational
masses of the neutron star residues would be too small 
(1.1--1.2 \msun) to 
explain the observations (which, however, could indeed be 
selection--biased).
Until other groups weigh in on this and credible 3--D 
calculations are
performed, these questions will remain open, but intriguing.  
The resolution
of these questions will be intimately coupled to the systematics 
of explosion
energy with progenitor mass and structure and there may yet be 
some surprises.
The progenitor structures themselves have not converged 
\cite{ww95,nh88,ba94}.
The binding energies of the mantles may well regulate the 
explosion and/or
determine its energy \cite{bhf95} and mass cut.

Whatever the final word, that the character of the explosion is 
multi-dimensional
seems robust.  Supernovae do not explode as spheres, but 
aspherically in plumes,
resembling cauliflower and brocolli more than oranges.  
Sato and collaborators
\cite{ssy93,sys93,sys94,ss94} even suggest that rapid rotation 
leads to aspherical neutrino emissions
that translate into jet--like neutrino--driven explosions.  
Be that as it may,
the exploration of many of the major questions of supernova theory 
and supernova explosions 
may now require multi--dimensional treatment.

\section{Neutrino Transport and Neutrino Opacities}

Though the new hydrodynamic issues are important, equal weight should 
be given
to the whole subject of neutrino transport and neutrino/matter 
interactions.
Core--collapse supernovae are the only context, apart from the 
big bang,
in which neutrinos are pivotal agents of dynamics and evolution.  
Rather than present a
comprehensive discussion of neutrino transport, I will touch on a few  
important topics that should be the focus of future investigations.

There has been a lot of effort in the last decade to understand the 
character
of lepton trapping and to derive the trapped lepton fraction 
($Y_{l}$) and the
entropy generated during infall \cite{bruenn85,bruenn89a,mb90}.  
A larger $Y_{l}$ makes
for a more energetic bounce and generates the shock wave further out 
in interior
mass \cite{bl83}.  This leaves less mass through which the shock must 
fight to
emerge.  With the viability of the direct mechanism at stake, 
this focus was
understandable and a great deal of insight was gained 
\cite{bruenn89b}.  However,
the best transport calculations reveal that $Y_{l}$ is far too 
low, neutrino
losses when the shock breaks out of the neutrinospheres are far 
too large, and
nuclear dissociation is far too debilitating for the prompt 
mechanism to be
salvaged \cite{bhf95,mb90}.  The shock stalls into accretion, 
as stated in \S 2,
and must be revitalized.  As Gerry Brown has mused, this is the 
``pause that
refreshes,'' yet the duration ($T_d$) of this delay to explosion 
(one hundred
milliseconds, seconds ?) has not been established.  The value of 
$T_d$ must
determine in part the residual neutron star mass, whether a black 
hole forms,
the ejected nucleosynthesis, the explosion energy, and in fact most 
of the
interesting questions that surround the supernova event.  There are 
many
indications that the delays in the calculations of BHF and HBHFC 
are too
short.  Not only is $T_d$ a function of progenitor structure, 
but it hinges
on the character of the emergent neutrino luminosity ($L_\nu$) 
and spectra
as well.  Therefore, what determines them and their evolution 
determines the
outcome of collapse.  However, because of the numerous feedbacks 
in the
radiation hydrodynamics of collapse and bounce, the consequences 
of various
interesting neutrino processes have at times been exaggerated.  
One thinks
immediately of $\nu-\bar{\nu}$ annilation into $e^+-e^-$ pairs, 
which is effective only
near the neutrinospheres, where heating and cooling processes are 
always 
dominated by the more mundane charged--current absorptions on 
nucleons.
The process $\nu+\bar\nu \rightarrow e^++e^-$ can not be pivotal 
in reigniting
the supernova explosion \cite{bmcv89}, but should be included for 
completeness.  Let us proceed to highlight various neutrino 
processes whose
study might indeed be profitable and whose character has yet to 
be fully 
delineated.

Mu and tau neutrinos ($\nu_\mu, \bar{\nu}_{\mu}, \nu_\tau, 
\bar{\nu}_{\tau}$, 
hereafter ``$\nu_\mu$'s'') collectively carry away most of the 
binding energy
(50\%--60\%) of the neutron star.  Hence, their effect on its 
thermal evolution
is crucial.  It is thought that neutrino--electron scattering 
and inverse
pair annihilation are the processes most responsible for the energy 
eqilibration of the $\nu_\mu$'s and their emergent spectra.  However,
credible calculations imply that the inverse of nucleon--nucleon 
bremsstrahlung ({\it e.g.}, $n + n \rightarrow n + n + \nu \bar{\nu}$) 
is
more important \cite{suzuki} in equilibrating the $\nu_\mu$'s.  This 
process has not heretofore been incorporated in supernova simulations.
Preliminary calculations imply that the emergent $\nu_\mu$ spectra 
are softened
by this effect.
This has consequences for neutrino nucleosynthesis, in particular, 
since
the relevant inelastic neutral--current processes are stiff functions 
of neutrino
energy \cite{h88,whhh90}.

The ``pinching'' of the neutrino spectra \cite{mb90} in flux--limited, 
multi--group
calculations has a similar effect, but calculations of Mezzacappa \& 
Bruenn \cite{mb93a,mb93b}
and Burrows, Hayes \& Pinto \cite{bhp}, solving in the former case 
the Boltzmann
equation and in the latter case the full velocity--dependent transport 
equation (without flux--limiter) indicate that the flux at higher 
neutrino energies may
be {\it higher} than seen in flux--limited calculations.  The 
hardness of the neutrino
spectrum is crucial not only in neutrino nucleosynthesis 
calculations, but
in the supernova mechanism itself, since the neutrino--matter coupling
(heating) rate is an increasing function of $\nu_e$ and $\bar{\nu}_e$ 
energy.  For a given 
$\nu_e$ and $\bar{\nu}_e$ luminosity and average neutrino energy, 
the heating 
rate may indeed be higher when calculated accurately.  All else 
being equal,
a 10\% -- 30\% increase in this heating rate (due to a more accurate 
treatment
of neutrino transport) may facilitate explosion.

There has been sporadic interest over the years in corrections to the 
neutrino/matter cross sections due to collective effects at modest 
densities
(\sgreat 10$^{12}$ gm cm$^{-3}$).  Corrections to neutrino--nucleus
Freedman scattering due to finite nuclear size (form factor) 
\cite{ts75,bml81},
ion--ion correlation \cite{lb91}, and electron screening 
\cite{los88,hor92}
have been and continue to be the subject of study. (Note that the 
screening correction
for single nucleon scattering has yet to be estimated.)  All these 
effects {\it decrease}
the $\nu - A$ cross section, the form factor at high energies and 
the other
effects at low energies.  A preliminary estimate of the cumulative 
effect
of these corrections on protoneutron star cooling \cite{lb91} 
indicates
that the neutrino luminosites in the first seconds are higher as a 
result and this
is germane to the neutrino--driven mechanism \cite{b87,janka93}.  
However,
equally interesting are the nucleon--blocking and Fermi--liquid 
corrections
to neutrino--nucleon scattering and absorption at high densities 
(\sgreat 10$^{14}$ gm cm$^{-3}$) \cite{ss79,ip82}.  In particular, 
the axial--vector coupling
constant, $g_A$, is renormalized to a lower value due, among other 
things,
to the $\Delta$ resonance \cite{brho}. This decreases 
the dominant neutrino--matter cross section by of order a factor
of two \cite{bl86} and, hence, increases the neutrino luminosity 
from the core 
on timescales of not tens of milliseconds, but hundreds of 
milliseconds to seconds.
With the prospects of higher long--term luminosities (due to weaker 
neutrino/matter
coupling at high densities) and harder emergent electron neutrino 
spectra (with the resulting greater 
neutrino heating rates exterior to the neutrinospheres),  many of 
the anticipated
improvements in the neutrino transport sector of supernova modeling 
favor explosion.
It is only when the neutrino physics is well
in hand that the ultimate role of hydrodynamic instabilities can be 
assessed properly.

\section{Nucleosynthetic Constraints: The Mass Cut and the r--process}

The study of the production and ejection of heavy elements in 
supernova
explosions has a long pedigree and is too large a subject to be more 
than
superficially addressed here \cite{bbfh57,thn90}.  It involves the 
proper calculation of the pre-supernova nested ``onion--skin'' 
structure of
freshly synthesized elements (with its dependence on convective 
burning
algorithms, thermonuclear rates, and electron capture rates), the 
explosive
processing of the inner zones of the ejecta, and the hydrodynamics 
of the
explosion itself.  The latter is poorly handled by those who take 
great pains
with the former.  Those who have focussed on the mechanism have paid 
insufficient attention to the nucleosynthetic consequences.  Clearly, 
the
two theoretical domains should be fused in future investigations.

SN1987A is a treasure trove of information on all aspects of the 
core--collapse 
supernova phenomenon, yet its bounty is still underutilized by 
supernova
modelers.  I will not illustrate this statement in detail within 
the narrow confines 
of this brief report, but will summarize a few useful derived 
constraints
on the hydrodynamics of supernova explosions.

As pointed out by Thielemann, Nomoto, \& Hashimoto \cite{tnh96}, the 
observation of \nuc{57}{Co} and \nuc{57}{Fe} in SN1987A at 
$\sim$1.5 times
the solar ratio with \nuc{56}{Fe} \cite{dan90,Vara,Kurf} and the 
stiff dependence
of $\left[^{57}Fe \over ^{56}Fe\right]$ on $Y_{\rm e}$ force the 
eventual mass cut
separating neutron star (or black hole) from the ejecta to be near 
$Y_{\rm e}\sim
0.496-0.498$.  Similar conclusions can be drawn from the solar and 
SN1987A abundance 
ratios of \nuc{56}{Fe} with the neutron--rich isotopes \nuc{58}{Fe}, 
\nuc{58}{Ni},
\nuc{60}{Ni}, and \nuc{61,62}{Ni}. (Stable nickel was indeed 
detected in SN1987A
via the infrared 6.634$\,\mu$m line \cite{witt}).  In the progenitor 
models
of Nomoto \& Hashimoto \cite{nh88} and Woosley \& Weaver \cite{ww95}, 
this
cut is almost always exterior to the iron core and is often close to 
the
inner oxygen zones.  For the more massive progenitor models of 
Woosley \& Weaver \cite{ww95}
(\sgreat\ 19 \msun), this demarcation line is exterior to 
$2 \times 10^3$ kilometers
and 1.7 \msun, while for those of Nomoto \& Hashimoto \cite{nh88}, 
it is closer 
in in mass.  Be that as it may, all the extant successful supernova 
calculations
(BHF, HBHFC, JM) eject too much material whose 
neutron--richness is inconsistent with observed nucleosynthesis 
(including the
``$N=50$'' ({\it e.g.}, $Sc, Y, Zr$) abundances).  This implies 
either that the delay to 
explosion is longer than calculated, that generically there is 
fall back of
0.1--0.3 \msun of envelope, that the progenitor  models are 
inaccurate, or some 
combination of all three. In addition, the yields of $Si$, 
$S$, and $Ca$ in the
models of Woosley \& Weaver \cite{ww95} are too large to explain 
halo star
\cite{mcwill,mcwill2} and SN1987A abundances, unless the mass cut 
is further out than
heretofore assumed.  The models of Nomoto \& Hashimoto \cite{nh88} 
don't have
this problem to the same degree.  There is also a hint in the SN1987A 
Ginga
X--ray data that in order to avoid excessive photoelectric absorption 
in the SN1987A
debris cloud, the $Si$, $S$, and $Ca$ yields must be smaller than in 
the 
Woosley \& Weaver \cite{ww95} models \cite{burvan95}.

What the nucleosynthetic constraints are collectively telling us about
the explosion is not yet clear in its entirety, but they embolden 
one to
speculate nevertheless.  One conclusion to be drawn is that if the 
prompt mechanism
obtains, there must be appreciable fallback.  Since fallback is a 
function
of mantle binding energy, which increases with progenitor mass 
\cite{bhf95,ww96},
the low--mass massive stars (\sles 15 \msun ?) can not explode 
by this mechanism.
Since this binding energy is correlated with the other shock killers 
(breakout
neutrino losses, photodissociation, accretion ram), the more massive 
progenitors
can not explode directly either.  The bounce shock must stall, as all 
the best
hydrodynamic models imply, leaving the delayed mechanism by default.

How long is the delay, $T_d$?  For ``low--mass'' progenitors, since 
there may not be
appreciable fallback, the delay must be sufficient that the shocked 
material
does not reach densities large enough to result in appreciable 
electron capture
and the ejection of matter with anomalous $Y_{\rm e}$ 's.  
Furthermore, since
the iron and ONeMg cores of such progenitors have the lowest 
$Y_{\rm e}$ 's
(\sles 0.43), none of this core material can be ejected.  
It must be accreted.
How long this takes depends upon the progenitor radial structure. 
(Care must be 
taken in estimating and quoting these delays to distinguish between 
total 
collapse time and time since bounce.)  For the highest mass
progenitors that still explode, fallback and/or a long delay due
to higher accretion \mdot 's may naturally bury neutron--rich matter. 
Furthermore, it may be that the explosion energy is set in some way 
by the envelope
binding energy, in which case the lighter progenitors may explode 
with a lower
energy that could be inadequate to forestall fallback. Hence, there
may be feedback that necessitates fallback for both light and heavy 
progenitors.
Whether this is true remains to be investigated.

It is thought that the r--process nuclei are produced in the winds 
that emerge
from the protoneutron star after the explosion commences 
\cite{wh92,qian,woos94,twj94}.
The onset of this supersonic wind phase is clearly manifest 
by the emergence of a second shock wave and a contact discontinuity 
\cite{bhf95} and is suppressed 
until the large pressures in the inner supernova ejecta abate due 
to expansion.
How long the wind is suppressed depends upon neutrino transport and 
neutrino
luminosities, the delay to explosion, and the progenitor structure.  
It is 
possible that the more massive progenitors, with their denser and 
more bound envelopes,
suppress the wind long enough and/or smoother it with fallback that 
they do
not yield r--process nuclei.  This would be consistent with the 
suggestion
by Mathews, Bazan, \& Cowan \cite{mbc92} that only low--mass 
massive stars ($\sim$10
\msun) produce the r--process isotopes seen in halo stars.  This 
idea should be
reinvestigated using the new data of Sneden {\it et al.} 
\cite{sned94},
Cowan {\it et al.} \cite{cowan96}, and McWilliam {\it et al.} 
\cite{mcwill,mcwill2}
and a better $\left[Fe \over H\right]$ versus age relation 
\cite{twarog}.
Something very important concerning supernova explosions lurks 
in the halo
abundance data.  However, extracting it may require a more 
sophisticated
galactic chemical evolution model than employed to date.

For those supernovae that yield r--process nuclei, a simple 
(simplistic ?)
scenario suggests itself. 1) The stalled shock is reenergized 
after material
with $Y_{\rm e}$ 's near 0.497--0.5 is accreted. 2) The mass cut 
is near the mass
interior to the shock when it is relaunched. 3) After the material 
interior to 
the mass cut is accreted onto the protoneutron star and the pressure 
around it
abates, a neutrino--driven wind with lower $Y_{\rm e}$ 's, determined
in part by $\nu_e$ and $\bar{\nu}_e$ absorption, emerges. 
This must happen
when the driving neutrino luminosities have decayed enough so that 
$\dot M_w$'s, and, hence, 
$\Delta M_w$ ($=\int^\infty_{t_i} \dot M_w dt$) are small. 
In particular, $\Delta M_w$
should be below $\sim\!10^{-4}$ to $\sim\!10^{-3}$ \msun.  A large 
driving $L_\nu$ might result
in the wind ejection of too much neutron--rich material.  However, 
there may
be enough leeway in $\Delta M_w$ to accomodate the production and 
ejection of 
``$N=50$'' isotopes and perhaps p--process
nuclei in the wind's first phases ($Y_{\rm e} \le 0.49$) (see also 
Hoffman {\it et al.}
\cite{hwfm96}).  The wind's last, high--entropy, phases eject the 
r--process
nuclei.  Whether this scenario or similar scenarios holds up 
hopefully will be 
tested by the next generation of supernova models.  Whatever 
scenario obtains,
the iron--peak abundances require a mass cut above 
$Y_{\rm e} = 0.497$, while
the ``$N=50$,'' p--process, and r--process nucleosynthesis require a
$Y_{\rm e}$ below this value in less than a few times 
10$^{-3}$ \msun .

\section{Pulsar Proper Motions, Natal Kicks and Gravitational 
Radiation}

A new pulsar distance scale \cite{tc93}, recent pulsar proper 
motion data \cite{hla93}, 
and the recognition that pulsar
surveys are biased towards low speeds \cite{ll94} imply that radio 
pulsars are a high-speed population. 
Mean three-dimensional galactic speeds of 450$\pm$90 km s$^{-1}$ 
have been estimated \cite{ll94}, with measured transverse speeds of 
individual pulsars ranging from
zero to $\sim$1500 km s$^{-1}$.   Impulsive mass loss in
a spherical supernova explosion that occurs in a binary can impart 
to the nascent
neutron star a substantial kick that reflects its progenitor's 
orbital speed 
\cite{ggo70,rs85}.   
However, theoretical studies of binary evolution through the 
supernova phase have difficulty reproducing
velocity distributions with the required mean and dispersion 
\cite{dc87,it95}.   
This implies that neutron stars receive an extra kick at or 
after birth.

In the past, an off-center (and rapidly rotating) magnetic 
dipole \cite{ht75} 
and anisotropic neutrino radiation
\cite{w87,ss94,c84,bk95} have been
invoked to accelerate neutron stars.
A 1\% net asymmetry in the neutrino radiation of a neutron 
star's binding energy
results in a $\sim$300 km s$^{-1}$ kick.
However, Burrows \& Hayes \cite{bh96} have recently demonstrated 
that if the
collapsing Chandrasekhar core is mildly asymmetrical,
the young neutron star can receive a large impulse during the 
explosion in which
it is born.
In those calculations, rocket-like mass motions, not neutrinos, 
dominated the recoil, which reached $\sim$530 km s$^{-1}$.
Such a speed is large,
but is only $\sim$2\% of that
of the supernova ejecta.  
This asymmetry/recoil
correlation seems generic.
However, whether such asymmetries are themselves generic has yet 
to be demonstrated.
Recent calculations of convection 
during shell oxygen and silicon burning \cite{ba94} 
and theoretical arguments \cite{dlg95} 
suggest that the initial density, velocity, and
composition asymmetries  
might indeed be interesting.  

The impulse delivered to the core depends upon the dipole moments 
of the angular distribution of both the 
envelope momentum and the neutrino luminosity.  The gravitational 
waveform depends upon the corresponding
quadrupole moments.  Curiously, using the standard quadrupole 
formula, Burrows \& Hayes \cite{bh96} derived that due to
the intense and anisotropic early neutrino burst, 
the neutrino contribution to the metric strain, $h^{TT}_{zz}$, 
can dominate 
during the early post-bounce epoch.    
This is true despite the fact 
that the neutrinos do not dominate the recoil and is a consequence 
of their relativistic nature.
The gravitational waves are radiated between 10 and 500 Hz.   
and h$^{TT}_{zz}$ does
not go to zero with time.  Hence, there may be ``memory''  
\cite{bt87} in the
gravitational waveform from a protoneutron star that is 
correlated with
its recoil and neutrino emissions.  
This memory is a distinctive
characteristic of asymmetric collapse and explosion.  Using 
\cite{a92}, one finds
that the 2'nd-generation LIGO might be able to detect a signal from
a core collapse anywhere in our galaxy.
Even without rotation \cite{msmk91}, an asymmetric collapse can 
result in appreciable gravitational wave emission.
The simultaneous detection of the neutrino and gravitational 
radiation signatures and of the recoil 
would provide direct information concerning supernova dynamics.

The kick mechanism suggested by Burrows \& Hayes \cite{bh96} is 
but one of several that researchers
are now exploring.  The group of W. Benz, M. Herant,
C. Fryer, and S. Colgate is exploring the possibility that 
asymmetrical Bondi-Hoyle accretion of fallback by the
young neutron star could result is asymmetrical neutrino emission 
and a ``neutrino
rocket.''  In order for this mechanism to work, on the order of 
0.1--0.2 \msun must fall back and the
resulting angular asymmetry in the mass accretion (and, hence, the 
neutrino emission) must be large.  
One of the virtues of this mechanism is that the motion of the young 
neutron star 
might thereby be unstable to acceleration, until the ram pressure 
balances the neutrino term at some hundreds
of kilometers per second.

\section{Conclusion}

What I have presented here is a collection of thoughts on the current 
status
of core--collapse supernova theory and its components.  I have not 
answered
the questions posed in the Introduction because they are the subject 
of ongoing 
investigations, or should be.  Most are as yet unresolved.
However, I have attempted to map out a viable research strategy 
for the field
that will resolve many of the ambiguities that remain.  It is 
important not only
to obtain credible computer explosions, but to discover the 
systematics with
progenitor mass and composition of 
the explosion energies, ejecta nucleosynthesis,
neutron star masses, explosion morphologies, and natal kicks, among 
other things.
Such has been the recent progress that this truly comprehensive 
theory 
seems
within our grasp in the next few years.

\end{document}